\documentclass[12pt]{article}
\usepackage{graphicx}
\usepackage{lscape}
\usepackage{citesort}
\usepackage{amssymb}
\usepackage{appendix}
\usepackage{multirow}
\usepackage[table]{xcolor}
\usepackage{colortbl}
\definecolor{lightgray}{gray}{0.9}

\usepackage{mathrsfs}
\begin{document}
\def\qq{\langle \bar q q \rangle}
\def\uu{\langle \bar u u \rangle}
\def\dd{\langle \bar d d \rangle}
\def\sp{\langle \bar s s \rangle}
\def\GG{\langle g_s^2 G^2 \rangle}
\def\Tr{\mbox{Tr}}
\def\figt#1#2#3{
        \begin{figure}
        $\left. \right.$
        \vspace*{-2cm}
        \begin{center}
        \includegraphics[width=10cm]{#1}
        \end{center}
        \vspace*{-0.2cm}
        \caption{#3}
        \label{#2}
        \end{figure}
    }

\def\figb#1#2#3{
        \begin{figure}
        $\left. \right.$
        \vspace*{-1cm}
        \begin{center}
        \includegraphics[width=10cm]{#1}
        \end{center}
        \vspace*{-0.2cm}
        \caption{#3}
        \label{#2}
        \end{figure}
                }

\def\ds{\displaystyle}
\def\beq{\begin{equation}}
\def\eeq{\end{equation}}
\def\bea{\begin{eqnarray}}
\def\eea{\end{eqnarray}}
\def\beeq{\begin{eqnarray}}
\def\eeeq{\end{eqnarray}}
\def\ve{\vert}
\def\vel{\left|}
\def\ver{\right|}
\def\nnb{\nonumber}
\def\ga{\left(}
\def\dr{\right)}
\def\aga{\left\{}
\def\adr{\right\}}
\def\lla{\left<}
\def\rra{\right>}
\def\rar{\rightarrow}
\def\lrar{\leftrightarrow}
\def\nnb{\nonumber}
\def\la{\langle}
\def\ra{\rangle}
\def\ba{\begin{array}}
\def\ea{\end{array}}
\def\tr{\mbox{Tr}}
\def\ssp{{\Sigma^{*+}}}
\def\sso{{\Sigma^{*0}}}
\def\ssm{{\Sigma^{*-}}}
\def\xis0{{\Xi^{*0}}}
\def\xism{{\Xi^{*-}}}
\def\qs{\la \bar s s \ra}
\def\qu{\la \bar u u \ra}
\def\qd{\la \bar d d \ra}
\def\qq{\la \bar q q \ra}
\def\gGgG{\la g^2 G^2 \ra}
\def\q{\gamma_5 \not\!q}
\def\x{\gamma_5 \not\!x}
\def\g5{\gamma_5}
\def\sb{S_Q^{cf}}
\def\sd{S_d^{be}}
\def\su{S_u^{ad}}
\def\sbp{{S}_Q^{'cf}}
\def\sdp{{S}_d^{'be}}
\def\sup{{S}_u^{'ad}}
\def\ssp{{S}_s^{'??}}

\def\sig{\sigma_{\mu \nu} \gamma_5 p^\mu q^\nu}
\def\fo{f_0(\frac{s_0}{M^2})}
\def\ffi{f_1(\frac{s_0}{M^2})}
\def\fii{f_2(\frac{s_0}{M^2})}
\def\O{{\cal O}}
\def\sl{{\Sigma^0 \Lambda}}
\def\es{\!\!\! &=& \!\!\!}
\def\ap{\!\!\! &\approx& \!\!\!}
\def\md{\!\!\!\! &\mid& \!\!\!\!}
\def\ar{&+& \!\!\!}
\def\ek{&-& \!\!\!}
\def\kek{\!\!\!&-& \!\!\!}
\def\cp{&\times& \!\!\!}
\def\se{\!\!\! &\simeq& \!\!\!}
\def\eqv{&\equiv& \!\!\!}
\def\kpm{&\pm& \!\!\!}
\def\kmp{&\mp& \!\!\!}
\def\mcdot{\!\cdot\!}
\def\erar{&\rightarrow&}
\def\olra{\stackrel{\leftrightarrow}}
\def\ola{\stackrel{\leftarrow}}
\def\ora{\stackrel{\rightarrow}}
% .........................................................

\def\simlt{\stackrel{<}{{}_\sim}}
\def\simgt{\stackrel{>}{{}_\sim}}

% .........................................................

\title{
         {\Large
                 {\bf
                   More about the $B$ and $D$ mesons in nuclear matter
                 }
         }
      }

\author{\vspace{1cm}\\
{\small  K. Azizi$^1$ \thanks {e-mail: kazizi@dogus.edu.tr}\,\,, N. Er$^2$ \thanks {e-mail: nuray@ibu.edu.tr}, H.
Sundu$^3$} \thanks {e-mail: hayriye.sundu@kocaeli.edu.tr} \\
{\small $^1$ Department of Physics, Do\u gu\c s University,
Ac{\i}badem-Kad{\i}k\"oy, 34722 \.{I}stanbul, Turkey}\\
{\small $^2$ Department of Physics, Abant {I}zzet Baysal University,
G\"olk\"oy Kamp\"us\"u, 14980 Bolu, Turkey}\\
{\small $^2$ Department of Physics, Kocaeli University,
 41380 Izmit, Turkey}\\
}
\date{}

\begin{titlepage}
\maketitle
\thispagestyle{empty}
\begin{abstract}
We calculate the shifts in  decay constants of  the pseudoscalar $B$ and $D$ mesons in
nuclear medium in the frame work of QCD sum rules. We write those shifts in terms of the $B-N$ and $D-N$  scattering lengths and an extra phenomenological parameter
entered to calculations. Computing an appreciate forward scattering correlation function, we derive the QCD sum rules for the $B-N$ and $D-N$  scattering lengths and the extra phenomenological parameter
in terms of various operators in nuclear medium.  We numerically find the values of the shifts in the  decay constants compared to their vacuum values. Using the sum rules obtained, we also
determine the shifts in the masses of these particles due to nuclear matter and  compare the results obtained  with the previous predictions in the
literature. 
\\

PACS number(s): 11.55.Hx, 21.65.Jk, 14.40.Lb, 14.40.Nd
\end{abstract}
\end{titlepage}

% The aim of this article
% Standard model and new physics
% Topcolor-assisted technicolor model

                                 %%%%%%%%%%%%%%%%%%%%%%%%%%%%%%%%%%%%%%%
       %%%%%%%%%%%%%%%%%%%%%%%%%%%%%%%%%%%%%%%             %%%%%%%%%%%%%%%%%%%%%%%%%%%%%%%%%%%%%%%%%%
                                 %%%%%%%%%%%%%%%%%%%%%%%%%%%%%%%%%%%%%%%                         
\section{Introduction} 
Study the in-medium properties of hadrons can help us not only better understand the perturbative and non-perturbative natures of QCD, but also can play crucial role in analyzing the results of heavy 
ion collision experiments as well as understanding  the internal structures of the dense astrophysical objects like neutron stars.
 From the experimental side, there have been a lot of experiments such as CEBAF and RHIC etc. focused on the study of
the properties of hadrons in nuclear medium. The FAIR  and CBM  Collaborations  intend to study the in-medium properties of different hadrons including the charmed mesons. The PANDA Collaboration
also aims to focus on the study of the properties of  hadrons in charm sector  \cite{Lutz,Friman,CBM,PANDA}.

 Along the experimental progresses, there are many theoretical works devoted to the study of the in-medium properties of hadrons. 
 The basic properties of the  nuclear matter are determined in \cite{Drukarev1}. Some
finite-density problems and the saturation properties of nuclear
matter are studied in \cite{DrukarevA,hatsuda,adami}.
 In series of papers \cite{RJF,XJ1,Nielsen}, the authors have studied the effects of nuclear matter
on the masses of the nucleons. In \cite{Koike}, the $\rho$, $\omega$ and $\phi$ mesons-nucleon scattering lengths and their mass shifts in nuclear medium are investigated via QCD sum rules. 
\cite{Hayashigaki} applies the same method to investigate the mass modification of $D$-meson at finite density. In \cite{Hilger}, the authors expand the work of \cite{Hayashigaki} to study the mass shift of also
 $B$ meson in nuclear matter. The in-medium mass  modification  of the  scalar charm meson is investigated in \cite{Hilger2}, which is then extended to include also the mass modification of the scalar $B_0$ meson 
in \cite{Wang2011}.
For some studies of mainly mass shifts for different hadrons in nuclear medium  see for instance
 \cite{TDC2,Cohen45,Cohen,Drukarevnucl,Hogaasen,Wang2012,Wang,Lee,Asakawa,Leinweber,EGDur,Yasui,Hatsuda1995,Leupold,Ruppert,Peters,Klingl,Thomas,Thomas2005,Rapp,Lutz2006,nuray}. 

In the present study, we extend the works of \cite{Hayashigaki,Hilger} to investigate the modifications in the decay constants of the pseudoscalar $B$ and $D$ mesons in the framework of QCD sum rules.
Considering contributions of various operators in nuclear medium, we calculate the appreciate forward scattering correlation function in hadronic and operator product expansion (OPE) sides in nuclear matter to obtain the QCD sum rules
for the $B-N$ and $D-N$  scattering lengths and an extra phenomenological parameter entering the expressions of the modifications in the decay constants of the mesons under consideration. To study  the electromagnetic
structures and strong interactions of these mesons with other hadrons existing in the medium as well as for investigation of the $B$ decays into the charmed $D$ meson, we need to know also the modifications in the
decay constants of these mesons due to nuclear medium besides the modifications in their mass.  Our results can be useful in this respect. The results of the present work  can also  be used 
in analyses of the 
data obtained via heavy ion collisions held at different experiments.

The outline of the paper is as follows. In next section, after deriving the expressions of the modifications in the  decay constants, we get the QCD sum rules for the $B-N$ and $D-N$ 
 scattering lengths and an extra phenomenological
parameter via calculating an appreciate forward scattering correlation function in terms of both the hadronic parameters and QCD degrees of freedom in nuclear matter. Last section is devoted to the numerical analysis of 
the sum rules, obtaining the working regions for the auxiliary parameters entering the sum rules and numerical results on the shifts in the  decay constants as well as the masses of the $B$ and $D$ mesons . We also compare the obtained results on the physical
quantities under consideration with the existing predictions in the literature.

\section{In-medium modifications of the  decay constants of the  $D$ and $B$ mesons via QCD sum rules}
In order to calculate the shifts in the  decay constants of $D$ and $B$ mesons in nuclear matter, we start with the following two-point correlation function which can be divided into the 
 vacuum  $\Pi_0 (q)$ and  the
static one-nucleon  $\Pi_N (q)$ parts in Fermi gas approximation for the nuclear matter.  The $\Pi_N (q)$ function can also  be approximated in the
linear density of the nuclear matter as
\cite{Drukarev,Hayashigaki}:
\begin{equation}\label{correilk}
\Pi(q)=i\int{d^4 xe^{iq\cdot x}\langle {\cal T}[J_{B[D]}
(x)J_{B[D]}^{\dag}(0)]\rangle_{\rho_N}}=\Pi_0(q)+\Pi_N(q)\simeq
\Pi_0(q)+\frac{\rho_N}{2M_N}T_N(q),
\end{equation}
where $ {\cal T}$ is the time ordering operator, $\rho_N$ is the density of the
nuclear matter,  $M_N$ is the mass of
the nucleon and $J_{B[D]}(x)$ denotes the interpolating current of the $B[D]$ meson. To find the shifts in the values of the  decay constants, we shall consider the forward
scattering amplitude $T_N(q)$ which can be written as
\begin{equation}\label{TN}
T_N(q_0=\omega,\bold{q})=i\int{d^4 xe^{iq\cdot x}\langle N(p)| {\cal T}[J_{B[D]}
(x)J_{B[D]}^{\dag}(0)]|N(p)\rangle},
\end{equation}
where $q^\mu=(\omega,\bold{q})$ is the four-momentum of the meson and
$|N(p)\rangle$ represents the
isospin and spin averaged static nucleon state which is normalized covariantly as $\langle
N(p)|N(p')\rangle=(2\pi)^32p_0 \delta^3(\bold{p}-\bold{p'})$
\cite{Hayashigaki,Wang2011}. The pseudoscalar $B[D]$-meson interpolating field is taken as
\begin{equation}\label{currentB}
J_{B[D]}(x)=\frac{\bar{u}(x)i\gamma_5 b[c](x)+\bar{b}[\bar{c}](x)i\gamma_5
u(x)}{2}.
\end{equation}
where $u(x)$, $b(x)$ and $c(x)$ are quark fields.  Note that in evaluating the $T_N(q)$ function we need to know the condensates $\langle{\cal O}_i\rangle_N$ which are related to the 
condensates $\langle{\cal O}_i\rangle_{\rho_N}$ via the following equation valid at relatively low density \cite{drukbey}:
\begin{equation}\label{sina}
 \langle{\cal O}_i\rangle_{\rho_N}=\langle{\cal O}_i\rangle_0+\frac{\rho_N}{2 M_N}\langle{\cal O}_i\rangle_N+o(\rho_N).
\end{equation}

In the following, our main goal is to evaluate the  forward
scattering amplitude to find the shifts in the  decay constants.
According to  the general philosophy of the method, we calculate this function via two different ways:
in the phenomenological or hadronic side using the hadronic parameters
and in  the OPE or theoretical side
in terms of QCD degrees of freedom. Equating these two
representations of the same function, we obtain  QCD sum rules for the shifts in the physical quantities under consideration.  To
suppress contributions of the higher states and continuum,  Borel transformation
and continuum subtraction are applied to both sides of the obtained sum rules.

\subsection{Hadronic Side}
The  forward scattering amplitude $T_N(\omega,\bold{q})$ 
is calculated in terms of the hadronic
parameters
 in the limit $\textbf{q}\rightarrow 0$, around
$\omega=m_{B[D]}$. Near the
pole position of the pseudoscalar meson, $T_N(\omega,0)$ is
related to the $\mathbf{T}$-matrix for the forward $B[D]-N$
scattering amplitude \cite{Koike}. The function $T_N(\omega,0)$ is
written  as the
following dispersion integrals \cite{Koike}:
\begin{eqnarray}\label{Tmatrix}
T_N(\omega,0)=\int^{+\infty}_{-\infty}du\frac{\rho(u,\textbf{q}=0)}{u-\omega-i\varepsilon}=\int_{0}^{\infty}du^2\frac{\rho(u,\textbf{q}=0)}{u^2-\omega^2},
\end{eqnarray}
where $\omega^2\neq $positive real number and the spin-averaged spectral density
$\rho(u,\textbf{q}=0)$ can be expressed in terms of the spin-averaged $B[D]-N$
scattering $\mathbf{T}$-matrix, decay constant and mass of the $B[D]$ meson as well as  phenomenological parameters $a$, $b$ and $c$ in the following way:
\begin{eqnarray}\label{rho}
\rho(u>0,\textbf{q}=0)&=&-\frac{f_{B[D]}^2m_{B[D]}^4}{\pi
m_{b[c]}^2}\mathbf{Im}\Big[\frac{\mathbf{T}_{B[D]N}(u,0)}
{(u^2-m_{B[D]}^2+i\varepsilon)^2}\Big]+...
\nonumber \\
&=&-\frac{f_{B[D]}^2m_{B[D]}^4}{\pi
m_{b[c]}^2}\Big\{\mathbf{Im}\frac{1}{(u^2-m_{B[D]}^2+i\varepsilon)^2}
\mathbf{Re}[\mathbf{T}_{B[D]N}(u,0)]
\nonumber \\
&+& \mathbf{Re}\frac{1}{(u^2-m_{B[D]}^2+i\varepsilon)^2}
\mathbf{Im}[\mathbf{T}_{B[D]N}(u,0)]\Big\}+...
 \\
&\equiv&a\frac{d}{du^2}\delta(u^2-m_{B[D]}^2)+b~\delta(u^2-m_{B[D]}^2)
+c~\delta(u^2-s_0),
\end{eqnarray}
where ... in Eq. (6) denotes the contribution of higher states and continuum which is not associated with the $B[D]-N$
scattering. It is equivalent to the third term in Eq. (7) which represents the scattering contribution in the continuum part of the  $B[D]$ current starting at the threshold $s_0$. 
Applying the Borel transformation and continuum subtraction  suppresses this contribution. Note that the first term proportional to the parameter $a$ in Eq.(7) denotes
the double-pole term and corresponds to the on-shell effect of the
$\mathbf{T}$-matrix. The second term proportional to the parameter $b$ in
Eq.(7) denotes the single-pole term and corresponds to the
off-shell effect of the $\mathbf{T}$-matrix.
The  phenomenological parameters $a$ and $b$ are found as
\begin{eqnarray}\label{ab}
a&=&-\frac{f_{B[D]}^2m_{B[D]}^4}{m_{b[c]}^2}\mathbf{Re}[\mathbf{T}_{B[D]N}(u,0)]
|_{u=m_{B[D]}}=-\frac{8\pi
f_{B[D]}^2m_{B[D]}^4(M_N+m_{B[D]})}{m^2_{b[c]}}a_{B[D]},
\nonumber \\
b&=&-\frac{f_{B[D]}^2m_{B[D]}^4}{m_{b[c]}^2}\frac{d}{du^2}\mathbf{Re}[
\mathbf{T}_{B[D]N}(u,0)]|_{u=m_{B[D]}},
\end{eqnarray}
where the parameter $a_{B[D]}$ is the $B[D]-N $ scattering length  \cite{Koike}. 
The decay constant
$f_{B[D]}$  of the pseudoscalar $B[D]$ meson is defined as
\begin{eqnarray}\label{mesonmatrix}
{\langle}0|J_{B[D]}(0)|B[D]{\rangle}=\frac{f_{B[D]}m_{B[D]}^2}{m_{b[c]}}.
\end{eqnarray}

Combining  Eqs.(7), (5), (2) and (1), we can relate the phenomenological parameters $a$ and $b$ extracted from the   forward scattering amplitude $T_N$
with the shifts in the mass and decay constant of the $B[D]$ meson as  \cite{Koike}:
\begin{eqnarray}\label{CorrelationFunc1}
\Pi^{HAD}(\omega,0)&\varpropto& \frac{F}{m_{B[D]}^2-\omega^2}+\frac{\rho_N}{2M_N}\Big\{\frac{a}{(m_{B[D]}^2-\omega^2)^2}+\frac{b}{m_{B[D]}^2-\omega^2}\Big\}\nonumber\\
&\simeq&\frac{F+\delta
F}{(m_{B[D]}^2+\Delta m_{B[D]}^2)-\omega^2}
\end{eqnarray}
where
\begin{eqnarray}\label{F}
F&=&\frac{f_{B[D]}^2m_{B[D]}^4}{m_{b[c]}^2},
\nonumber \\
\delta F&=&\frac{\rho_N}{2M_N}b,
\nonumber \\
\Delta m_{B[D]}^2&=&-\frac{\rho_N}{2M_NF}a.
\end{eqnarray}
Using the modified mass in nuclear matter, $m_{B[D]}^*=m_{B[D]}+\delta
m_{B[D]}=\sqrt{m_{B[D]}^2+\Delta m_{B[D]}^2}$, the mass shift of
B[D] meson is obtained as:
\begin{eqnarray}\label{massshift}
\delta m_{B[D]}=2\pi\frac{M_N+m_{B[D]}}{M_N m_{B[D]}}\rho_N
a_{B[D]}.
\end{eqnarray}
From  Eq.(\ref{F}), the shift in 
decay constant of the $B[D]$ meson is also obtained as
\begin{eqnarray}\label{lepdecconst}
\delta
f_{B[D]}=\frac{m_{b[c]}^2}{2f_{B[D]}m_{B[D]}^4}\Big(\frac{\rho_N}{2M_N}b-\frac{
4f_{B[D]}^2
m_{B[D]}^3}{m^2_{b[c]}}\delta m_{B[D]}\Big).
\end{eqnarray}

As it is clear from the above relations, to find the shifts in the  mass and decay constant, we need to calculate the phenomenological  parameters
$a$ and  $b$ using the forward scattering amplitude calculated both in hadronic and OPE sides.

In the low energy
limit $\omega\rightarrow 0$, the  $T^{HAD}_N(\omega,0)$ is
equivalent to the Born term $T_N^{Born}(\omega,0)$. Hence, the forward scattering amplitude in hadronic side can be written as
\begin{eqnarray}\label{case1}
T^{HAD}_N(\omega,0)=T_N^{Born}(\omega,0)+\frac{a}{(m_{B[D]}^2-\omega^2)^2}+
\frac{b}{m_{B[D]}^2-\omega^2}+\frac{c}{s_0^2-\omega^2},
\end{eqnarray}
with the condition
\begin{eqnarray}\label{cond1}
\frac{a}{m_{B[D]}^4}+\frac{b}{m_{B[D]}^2}+\frac{c}{s_0}=0.
\end{eqnarray}

The Born term can be determined by the Born diagrams at the tree
level \cite{Koike,Hayashigaki}. To calculate it, we consider the contributions of the baryons $\Lambda_{b[c]}$ and $\Sigma_{b[c]}$ in the medium produced by the interaction of B[D] with the nucleon, i.e.
\begin{eqnarray}\label{DNInt}
B^-(b\overline{u})+p(uud)~or~n(udd)&\rightarrow&
\Lambda_b^0(udb)~or~\Sigma_b^-(ddb),
\nonumber \\
D^0(c\overline{u})+p(uud)~or~n(udd)&\rightarrow& \Lambda_c^+,
\Sigma_c^+(udc)~or~\Sigma_c^0(ddc)
\end{eqnarray}
The Born term $T_N^{Born}(\omega,0)$ is
obtained as \cite{Hayashigaki}:
\begin{eqnarray}\label{TBorn}
T^{Born}(\omega,
0)=\frac{2M_N(M_N+M_\mathcal{B})m_{B[D]}^4f_{B[D]}^2}{[\omega^2-(M_N+M_\mathcal{B}
)^2](\omega^2-m_{B[D]}^2)^2(m_u+m_{b[c]})^2}
g_{NB[D]\mathcal{B}}^2(\omega^2).
\end{eqnarray}
where $\mathcal{B}$ denotes the $\Lambda_{b[c]}$ or $\Sigma_{b[c]}$ baryon and $g_{NB[D]\mathcal{B}}(\omega^2)$ is the strong coupling constant among the $B[D]$ meson, nucleon and $\mathcal{B}$ baryon.

After Borel transformation and using the quark-hadron duality assumption, the hadronic side of the
current-nucleon forward scattering amplitude is
obtained as (see also \cite{Hayashigaki}):
\begin{eqnarray}\label{Phys}
\widehat{\textbf{B}}T_N^{HAD}&=&a\Big(\frac{1}{M^2}e^{-m_{B[D]}^2/M^2}-\frac{
s_0}{m_{B[D]}^4}
e^{-s_0/M^2}\Big)+b\Big(e^{-m_{B[D]}^2/M^2}-\frac{s_0}{m_{B[D]}^2}
e^{-s_0/M^2}\Big)
\nonumber \\
&+&\frac{2f_{B[D]}^2m_{B[D]}^4M_N(M_N+M_\mathcal{B})}{[(M_N+M_\mathcal{B})^2-m_{
B[D]}^2]
(m_u+m_{b[c]})^2}g_{NB[D]\mathcal{B}}^2
\nonumber \\
&\times& \Big[-\frac{e^{-(M_N+M_\mathcal{B})^2/M^2}}
{(M_N+M_\mathcal{B})^2-m_{B[D]}^2}+\Big(\frac{1}{(M_N+M_\mathcal{B})^2-m_{B[D]}
^2}-\frac{1}{M^2}\Big)
e^{-m_{B[D]}^2/M^2}\Big].
\nonumber \\
\end{eqnarray}

\subsection{OPE side}
The OPE side of the forward scattering amplitude is obtained via inserting the  explicit form of the interpolating current $J_{B[D]}$ into Eq. (\ref{TN}). After contracting out all
quark pairs via Wick's theorem, we get
\begin{eqnarray}\label{corre2}
T_N^{OPE}&=&\frac{i}{4} \int d^4 x e^{iq.x}\Bigg\langle N(p)\Bigg
|Tr\Big[S_Q(-x)\gamma_5 S_u(x)\gamma_5+S_u(-x)\gamma_5 S_Q(x)\gamma_5\Big]\Bigg|
N(p)\Bigg\rangle,\nonumber\\
\end{eqnarray}
where $S_u$ is light quark and $S_Q$ with $Q=b$ or $c$ is the  heavy quark propagator. The light quark
propagator   in the fixed-point gauge at  nuclear medium is given by
\cite{Cohen,Reinders}:
\begin{eqnarray}\label{propagatorlight}
 S_{u}^{ij}(x)&\equiv& \langle
 N(p)|T[q^i
(x)\bar{q}^j(0)]|N(p)\rangle\nonumber \\
&=&
\frac{i}{2\pi^2}\delta^{ij}\frac{1}{(x^2)^2}\not\!x
-\frac{m_q }{ 4\pi^2} \delta^ { ij } \frac { 1}{x^2} \nonumber \\
&+&
\chi^i_q(x)\bar{\chi}^j_q(0)-\frac{ig_s}{32\pi^2}F_{\mu\nu}^A(0)t^{ij,A
}\frac{1}{x^2}[\not\!x\sigma^{\mu\nu}+\sigma^{\mu\nu}\not\!x]+...,
\end{eqnarray}
where $\chi^i_q$ and $\bar{\chi}^j_q$
are the Grassmann background quark fields and  $F_{\mu\nu}^A$ is classical
background gluon field. The first and second terms in the above equation stand for free or perturbative part, and the third and forth
terms denote  the non-perturbative part or contributions due to the background quark and gluon fields. The heavy quark propagator is taken as
\begin{eqnarray}\label{propagatorheavy}
 S_{Q}^{ij}(x)&\equiv& \langle N(p)|T[Q^i
(x)\bar{Q}^j(0)]|N(p)\rangle \nonumber \\
&=&
\frac{i}{(2\pi)^4}\int d^4 k e^{(-ik\cdot x)}\Bigg\{
\frac{\delta^{ij}}{\not\!k-m_Q}-\frac{g_s
G^n_{\alpha\beta}t_{ij}^n}{4}\frac{\sigma^{\alpha\beta}
(\not\!k+m_Q)+(\not\!k+m_Q)\sigma^{\alpha\beta}}{(k^2-m_Q^2)^2}\nonumber
\\&&+\frac{ \delta_{ij} \langle g_s^2 GG\rangle }{12}\frac{m_Qk^2+m_Q^2
\not\!k}{(k^2-m_Q^2)^4}+... \Bigg\} .
\end{eqnarray}

The next step is to use the light and heavy quark propagators in Eq. (\ref{corre2}). As we deal only with the shifts in the mass and decay constant compared to their vacuum values, 
it is enough to consider only  the terms having non-perturbative effects. To go further, we need to define 
the products of the Grassmann background quark fields and classical background
gluon fields in terms of
the ground-state matrix elements of the corresponding quark and gluon operators at nuclear medium
\cite{Cohen},
\begin{eqnarray}\label{kaboli}
&&\chi_{i\alpha}^{q}(x)\bar{\chi}_{j\beta}^{q}(0)=\langle
q_{i\alpha}(x)\bar{q}_{ j\beta}(0)\rangle_N, ~~~~~~~
F_{\kappa\lambda}^{A}F_{\mu\nu}^{B}=\langle
G_{\kappa\lambda}^{A}G_{\mu\nu}^{B}\rangle_N, \nonumber \\
&&\chi_{i\alpha}^{q}\bar{\chi}_{j\beta}^{q}F_{\mu\nu}^{A}=\langle
q_{i\alpha}\bar{q}_{ j\beta}G_{\mu\nu}^{A}\rangle_N, 
~~~~~~~~\chi_{i\alpha}^{q}\bar{\chi}_{j\beta}^{q}\chi_{k\gamma}^{q}\bar
{\chi}_{l\delta}^{q}=\langle
q_{i\alpha}\bar{q}_{j\beta}
q_{k\gamma}\bar{q}_{l\delta}\rangle_N.
\end{eqnarray}
 The matrix elements $\langle
q_{i\alpha}(x)\bar{q}_{j\beta}(0)\rangle_N$ and $\langle
g_{s}q_{i\alpha}\bar{q}_{j\beta}G_{\mu\nu}^{A}\rangle_N$ are defined as
\cite{Cohen} 
\begin{eqnarray} \label{ }
\langle
q_{i\alpha}(x)\bar{q}_{j\beta}(0)\rangle_N&=&-\frac{\delta_{ij}}{12}\Bigg
[\Bigg(\langle\bar{q}q\rangle_N+x^{\mu}\langle\bar{q}D_{\mu}q\rangle_N
+\frac{1}{2}x^{\mu}x^{\nu}\langle\bar{q}D_{\mu}D_{\nu}q\rangle_N
+...\Bigg)\delta_{\alpha\beta}\nonumber \\
&&+\Bigg(\langle\bar{q}\gamma_{\lambda}q\rangle_N+x^{\mu}\langle\bar{q}
\gamma_{\lambda}D_{\mu} q\rangle_N
+\frac{1}{2}x^{\mu}x^{\nu}\langle\bar{q}\gamma_{\lambda}D_{\mu}D_{\nu}
q\rangle_N
+...\Bigg)\gamma^{\lambda}_{\alpha\beta} \Bigg],\nonumber \\
\end{eqnarray}
 and
\begin{eqnarray} \label{ }
\langle
g_{s}q_{i\alpha}\bar{q}_{j\beta}G_{\mu\nu}^{A}\rangle_N&=&-\frac{t_{
ij}^{A
}}{96}\Bigg\{\langle g_{s}\bar{q}\sigma\cdot
Gq\rangle_N\Bigg[\sigma_{\mu\nu}+i(u_{\mu}\gamma_{\nu}-u_{\nu}\gamma_{\mu
})
\!\not\! {u}\Bigg]_{\alpha\beta} \nonumber \\
&&+\langle g_{s}\bar{q}\!\not\! {u}\sigma\cdot
Gq\rangle_N\Bigg[\sigma_{\mu\nu}\!\not\!
{u}+i(u_{\mu}\gamma_{\nu}-u_{\nu}\gamma_{\mu}
)\Bigg]_{\alpha\beta} \nonumber \\
&&-4\Bigg(\langle\bar{q}u\cdot D u\cdot D q\rangle_N+im_{q}\langle\bar{q}
\!\not\! {u}u\cdot D q\rangle_N\Bigg) \nonumber \\
&&\times\Bigg[\sigma_{\mu\nu}+2i(u_{\mu}\gamma_{\nu}-u_{\nu}\gamma_{\mu}
)\!\not\! {u}\Bigg]_{\alpha\beta}\Bigg\},
\end{eqnarray}
where  $D_\mu=\frac{1}{2}(\gamma_\mu
\!\not\!{D}+\!\not\!{D}\gamma_\mu)$.
The matrix element of the four-dimension gluon condensate is also defined as
\begin{equation}
 \langle
G_{\kappa\lambda}^{A}G_{\mu\nu}^{B}\rangle_N=\frac{\delta^{AB}}{96}
\Bigg[
\langle
G^{2}\rangle_N(g_{\kappa\mu}g_{\lambda\nu}-g_{\kappa\nu}g_{\lambda\mu}
)+O(\langle
\textbf{E}^{2}+\textbf{B}^{2}\rangle_N)\Bigg],
\end{equation}
where we neglect the last term in this equation because of its small
contribution. We also ignore from the four-quark condensate contributions in Eq. (\ref{kaboli}).

Various condensates appear in calculations are defined in terms of the four-velocity $u^\mu$ of the nuclear medium  as \cite{Cohen,XJ1}
\begin{eqnarray} \label{ }
\langle\bar{q}\gamma_{\mu}q\rangle_N&=&\langle\bar{q}\!\not\!{u}q\rangle_N
u_{\mu} , \\
\langle\bar{q}D_{\mu}q\rangle_N&=&\langle\bar{q}u\cdot D
q\rangle_N
u_{\mu}=-im_{q}\langle\bar{q}\!\not\!{u}q\rangle_N
u_{\mu}  ,\\
\langle\bar{q}\gamma_{\mu}D_{\nu}q\rangle_N&=&\frac{4}{3}\langle\bar{q}
\!\not\! {u}u\cdot D q\rangle_N(u_{\mu}u_{\nu}-\frac{1}{4}g_{\mu\nu}) 
+\frac{i}{3}m_{q} \langle\bar{q}q\rangle_N(u_{\mu}u_{\nu}-g_{\mu\nu}),
\\
\langle\bar{q}D_{\mu}D_{\nu}q\rangle_N&=&\frac{4}{3}\langle\bar{q}
u\cdot D u\cdot D q\rangle_N(u_{\mu}u_{\nu}-\frac{1}{4}g_{\mu\nu}) 
-\frac{1}{6} \langle
g_{s}\bar{q}\sigma\cdot Gq\rangle_N(u_{\mu}u_{\nu}-g_{\mu\nu}) ,\\
\langle\bar{q}\gamma_{\lambda}D_{\mu}D_{\nu}q\rangle_N&=&2\langle\bar{q}
\!\not\! {u}u\cdot D u\cdot D
q\rangle_N\Bigg[u_{\lambda}u_{\mu}u_{\nu} -\frac{1}{6} 
(u_{\lambda}g_{\mu\nu}+u_{\mu}g_{\lambda\nu}+u_{\nu}g_{\lambda\mu})\Bigg]
\nonumber\\
&&-\frac{1}{6} \langle
g_{s}\bar{q}\!\not\! {u}\sigma\cdot
Gq\rangle_N(u_{\lambda}u_{\mu}u_{\nu}-u_{\lambda}g_{\mu\nu}),
\end{eqnarray}
where the  equation of motion have been used and terms $\textit{O}(m^2_q)$  have
been neglected due to their very small contributions \cite{Cohen}.

Making use of all  above equations,  the  OPE side of the $T_N$ function in the rest frame of the nuclear matter  in Borel scheme is obtained as
\begin{eqnarray}\label{QCDside}
\widehat{\textbf{B}}T^{OPE}_N&=&\frac{1}{3}\frac{
e^{-m_Q^2/M^2}}{M^4}\Bigg\{-m_Q
\Big(-2m_Q^2+M^2+2p_0^2\Big)\langle \bar{q}g_{s}\sigma
Gq\rangle_N\nonumber \\
&&-4m_Q \Big(m_Q^2-2M^2+4p_0^2\Big)\langle\bar{q}D_{0}D_{0}
q\rangle_N \nonumber \\
&&+4M^2\Big(-m_Q^2+M^2+4p_0^2\Big)\langle
q^{\dag}iD_{0}
q\rangle_N \nonumber \\
&&+2M^2\Bigg[2m_Q^2m_u-3m_QM^2+m_u\Big(M^2-2p_0^2\Big)\Bigg]\langle
\bar{q}q\rangle_N\Bigg\} \nonumber \\
&&+ \frac{1}{12\pi^2}\langle
g_{s}^{2}G^{2}\rangle_N\int_0^{\infty}d\alpha
\frac{e^{m_Q^2/(4\alpha-M^2)}m_Q}{\Big(4\alpha-M^2\Big)^4} \Bigg\{16\alpha^2
\Big(m_Q+3m_u\Big) \nonumber \\
&&+M^2\Big(-m_Q^2m_u+3m_Q M^2+3m_u M^2\Big) \nonumber \\
&&-4\alpha\Big(m_Q^3-m_Q^2 m_u +4m_Q M^2+6m_u M^2\Big) \Bigg\}
\theta\Bigg[\frac{1}{-4\alpha+M^2}\Bigg] \nonumber \\
&&-\frac{m_Q
e^{-m_Q^2/M^2}}{M^2}\langle \bar{q}g_{s}\sigma
Gq\rangle_N.
\end{eqnarray}

\subsection{QCD sum rules for the phenomenological parameters $a$ and $b$}
In this subsection, the Borel transformed  hadronic and OPE sides of the $T_N$ function are equated to find QCD sum rules for the 
parameters $a$ and $b$, i.e.,
\begin{equation}\label{findab1}
\widehat{\textbf{B}}T_N^{HAD}=\widehat{\textbf{B}}T_N^{OPE}.
\end{equation}
As we have two unknowns, we need  one more equation which is find
applying derivative with respect to $\frac{1}{M^2}$ to both sides of 
Eq. (\ref{findab1})
\begin{equation}\label{findab2}
\frac{\partial}{\partial(1/M^2)}\Big\{\widehat{\textbf{B}}T_N^{HAD}\Big\}=  \frac{\partial}{ \partial(1/M^2)}\Big\{\widehat{\textbf{B}}T_N^ { OPE}\Big\}.
\end{equation}
By simultaneous solving of equations  (\ref{findab1}) and  (\ref{findab2}), we
obtain the following sum rules for the parameters $a$ and $b$:
\begin{eqnarray}\label{a}
a&=&\frac{f_{B[D]}^2g_{NB[D]\mathcal{B}}^2}{\Delta~\Delta^{\prime}(m_u+m_{b[c]}
)^2}
\Bigg\{-2m_{B[D]}^4M_N(M_N+M_\mathcal{B})\exp\Big[-\frac{m_{B[D]}^2+(M_\mathcal{
B}+M_N)^2}{M^2}\Big]
\nonumber \\
&+&\frac{2s_0m_{B[D]}^2M_N
(M_N+M_\mathcal{B})\Big((M_N+M_\mathcal{B})^2-s_0\Big)}{\Delta'}\exp
\Big[-\frac{(M_\mathcal{B}+M_N)^2+s_0}{M^2}\Big]
\nonumber \\
&+&
2m_{B[D]}^4M_N(M_N+M_\mathcal{B})\exp\Big[-\frac{2m_{B[D]}^2}{M^2}\Big]
-\frac{2s_0m_{B[D]}^2M_N(M_N+M_\mathcal{B})}{\Delta' M^2}
\Big[m_{B[D]}^4+M_N^2M^2
\nonumber \\
&+& M_N^2s_0-M^2s_0-m_{B[D]}^2(M_N^2+s_0)+(s_0+M^2
-m_{B[D]}^2)(M_\mathcal{B}^2+2M_NM_\mathcal{B})\Big]
\nonumber \\
&\times& \exp\Big[-\frac{m_{B[D]}^2+s_0}{M^2}\Big]
\Bigg\}+\frac{1}
{\Delta m_{B[D]}^2}\Bigg\{m_{B[D]}^2\Big[m_{B[D]}^2
\widehat{\textbf{B}}T_N^ { OPE}+\Big(\frac{d}{d\frac{1}{M^2}}\widehat{\textbf
{B}}T_N^ { OPE}
\Big)\Big]
\nonumber \\
&\times& \exp\Big[-\frac{m_{B[D]}^2}{M^2}\Big]-s_0\Big[s_0
\widehat{\textbf{B}}T_N^ { OPE}+\Big(\frac{d}{d\frac{1}{M^2}}\widehat{\textbf
{B}}T_N^ { OPE}\Big)
\Big]\exp\Big[-\frac{s_0}{M^2}\Big]\Bigg\},
\end{eqnarray}
\begin{eqnarray}\label{b}
b&=&\frac{f_{B[D]}^2g_{NB[D]\mathcal{B}}^2}{\Delta^{\prime}\Delta\Phi}\Bigg\{
-\frac{2M_Ns_0(M_N+M_\mathcal{B})}{\Delta^{\prime}M^2(m_u+m_{b[c]})^2}\Bigg[
M_N^2 M^2+M_\mathcal{B}^2(m_{B[D]}^2+M^2)+2M_\mathcal{B}
M_N(m_{B[D]}^2+M^2)
\nonumber \\
&+&m_{B[D]}^2(M_N^2-M^2-s_0)\Bigg]\exp\Big[-\frac{m_{B[D]}^2+\Big((M_N+M_\mathcal{B})^2
+s_0\Big)}{M^2}
\Big]+\frac{2m_{B[D]}^4M_N(M_N+M_\mathcal{B})}{M^2(m_u+m_{b[c]})^2}
\nonumber \\
&\times&
\exp\Big[-\frac{2m_{B[D]}^2+(M_N+M_\mathcal{B})^2}{M^2}\Big]+\frac{2s_0^2M_N
(M_N+M_\mathcal{B})(M_\mathcal{B}^2+2M_\mathcal{B}M_N+M_N^2-s_0)}
{\Delta^{\prime}m_{B[D]}^2(m_u+m_{b[c]})^2}
\nonumber \\
&\times&
\exp\Big[-\frac{(M_N+M_\mathcal{B})^2+2s_0}{M^2}\Big]+\frac{2s_0m_{B[D]}^2M_N
(M_N+M_\mathcal{B})(m_{B[D]}^2-s_0)}{\Delta^{\prime}M^2(m_u+m_{b[c]})^2}
\exp\Big[-\frac{2m_{B[D]}^2+s_0}{M^2}\Big]
\nonumber \\
&-&\frac{2s_0^2M_N(M_N+M_\mathcal{B})(m_{B[D]}^2-s_0)}{\Delta^{\prime}m_{B[D]}
^2(m_u+m_{b[c]})^2}
\exp\Big[-\frac{m_{B[D]}^2+2s_0}{M^2}\Big]-\frac{2m_{B[D]}^4M_N(M_N+M_\mathcal{B
})}{M^2(m_u+m_{b[c]})^2}
\nonumber \\
&\times&
\exp\Big[-\frac{3m_{B[D]}^2}{M^2}\Big]+\frac{2s_0M_N(M_N+M_\mathcal{B})
\Big(M^4+m_{B[D]}^2(M^2-m_{B[D]}^2+s_0)\Big)}{M^4(m_u+m_{b[c]})^2}
\nonumber \\
&\times&
\exp\Big[-\frac{2m_{B[D]}^2+s_0}{M^2}\Big]+\frac{2s_0^2M_N(M_N+M_\mathcal{B})
(m_{B[D]}^2-M^2-s_0)}{M^2m_{B[D]}^2(m_u+m_{b[c]})^2}\exp\Big[-\frac{m_{B[D]}
^2+2s_0}{M^2}\Big]
\nonumber \\
&+&\frac{2\Delta
m_{B[D]}^4M_N(M_N+M_\mathcal{B})(\Delta^{\prime}-M^2)}{\Delta^{\prime}M^2(m_u+m_
{b[c]})^2}
\exp\Big[-\frac{m_{B[D]}^2}{M^2}\Big]+\frac{2\Delta
m_{B[D]}^4M_N(M_N+M_\mathcal{B})}{\Delta^{\prime}(m_u+m_{b[c]})^2}
\nonumber \\
&\times& \exp\Big[-\frac{(M_N+M_\mathcal{B})^2}{M^2}\Big] \Bigg\}+
\frac{\widehat{\textbf{B}}T_N^ { OPE}}{\Phi \Delta
}\Bigg\{\Delta-\frac{m_{B[D]}^2}{M^2}\exp\Big[-\frac{2m_{B[D]}^2}{M^2}\Big]
+\frac{s_0(M^2+s_0)}{M^2m_{B[D]}^2}
\nonumber \\
&\times&
\exp\Big[-\frac{m_{B[D]}^2+s_0}{M^2}\Big]-\frac{s_0^3}{m_{B[D]}^6}
\exp\Big[-\frac{2s_0}{M^2}\Big]\Bigg\}+\frac{1}{\Delta\Phi}
\Big(\frac{d}{d\frac{1}{M^2}}\widehat{\textbf{B}}T_N^ { OPE}\Big)\Bigg\{\frac{
s_0(M^2+m_{B[D]}^2)}
{M^2m_{B[D]}^4}
\nonumber \\
&\times& \exp\Big[-\frac{m_{B[D]}^2+s_0}{M^2}\Big]-\frac{1}{M^2}
 \exp\Big[-\frac{2m_{B[D]}^2}{M^2}\Big]-\frac{s_0^2}{m_{B[D]}^6}
 \exp\Big[-\frac{2s_0}{M^2}\Big] \Bigg\},
\end{eqnarray}
where
\begin{eqnarray}\label{Delta}
\Delta^{\prime}&=&(M_\mathcal{B}+M_N)^2-m_{B[D]}^2
\nonumber \\
\Delta&=&\exp\Big[-\frac{2m_{B[D]}^2}{M^2}\Big]-\frac{2s_0
}{m_{B[D]}^2}\exp\Big[-\frac{m_{B[D]}^2+s_0}{M^2}\Big]+\frac{s_0
}{M^2}\exp\Big[-\frac{m_{B[D]}^2+s_0}{M^2}\Big]
\nonumber \\
&+& \frac{s_0^2 }{m_{B[D]}^4}
\exp\Big[-\frac{m_{B[D]}^2+s_0}{M^2}\Big] -\frac{s_0^2
}{m_{B[D]}^2M^2}\exp\Big[-\frac{m_{B[D]}^2+s_0}{M^2}\Big]
\nonumber \\
\Phi&=&\exp\Big[-\frac{m_{B[D]}^2}{M^2}\Big]-\frac{s_0}{m_{B[D]}^2}
\exp\Big[-\frac{s_0}{M^2}\Big].
\end{eqnarray}

\section{Numerical results and discussion}
In order to numerically analyze the QCD sum rules obtained in the previous section, we need to know the numerical values of the $\langle{\cal O}_i\rangle_N$ condensates. As we deal only with the shifts in the 
physical quantities under consideration with respect to their vacuum values, we set the vacuum condensates $\langle{\cal O}_i\rangle_0$ to zero in Eq. (\ref{sina}) and find the values of the $\langle{\cal O}_i\rangle_N$ condensates
in terms of the condensates $ \langle{\cal O}_i\rangle_{\rho_N}$, i.e. $\langle{\cal O}_i\rangle_N=\frac{2 M_N}{\rho_N} \langle{\cal O}_i\rangle_{\rho_N}$.  Using this relation and the values of condensates 
$ \langle{\cal O}_i\rangle_{\rho_N}$ presented in   \cite{Cohen,Nielsen,XJ1,Cohen45} we find the values of the condensates $\langle{\cal O}_i\rangle_N$ as depicted in table 1 (see also \cite{Wang2011}).  To proceed further
in numerical analysis, we also need the values of 
 some other input parameters like quark masses which are also presented in table 1. Note that,   in the present study, we use the quark masses in $\overline{MS}$ scheme. 
%As we also previously said, we assume that the linear density approximation is valid at the low nuclear density, i.e.,
%$ \langle{\cal O}_i\rangle_{\rho_N}=\langle{\cal O}_i\rangle_0+\frac{\rho_N}{2 M_N}\langle{\cal O}_i\rangle_N$. Since we deal with the shifts in the physical quantities under consideration with respect to their vacuum 
%values, we set the vacuum values of 
%the operators to zero and find the values of the condensates $\langle{\cal O}_i\rangle_N$ from this relation which are used in our numerical analysis.

Besides these input parameters, the sum rules for the parameters $a$ and $b$ contain two auxiliary objects, namely
the Borel mass parameter $M^2$ and continuum threshold $s_0$.
According to the general philosophy of the method used, the physical quantities should be independent of these auxiliary objects. Hence, we should look for ``working regions'' of these parameters such that at these
regions, the physical quantities have weak dependence on $M^2$ and $s_0$.
Our numerical calculations show that in the intervals $25~GeV^2 \leq M^2
\leq 40~GeV^2$ and $4~GeV^2 \leq M^2 \leq 8~GeV^2$ respectively in the $B$ and
$D$ channels, the dependence of the shifts in the physical quantities are weak. Also, we see that in the intervals
$34~GeV^2\leq s_0 \leq 38~GeV^2$ and $5.6~GeV^2\leq
s_0 \leq 6.4~GeV^2$ respectively for the $B$ and
$D$ mesons, the results demonstrate weak dependence on the continuum threshold.

\begin{table}[ht!]
\centering
\rowcolors{1}{lightgray}{white}
\begin{tabular}{cc}
\hline \hline
    Parameters   &  Values  
           \\
\hline \hline
$p_{0}^{B[D]}   $          &  $5.279 [1.870]  $ $GeV$      \\
%$ m_{u}   $          &  $5  $ $MeV$       \\
%$ m_{d}   $          &  $7  $ $MeV$       \\
%$ m_{b[c]}   $          &  $4.8[1.46]  $ $GeV$       \\
$ m_{b[c]}   $          &  $4.18[1.275]  $ $GeV$       \\
$ M_{N}   $          &  $0.938  $ $GeV$       \\
$ M_{\mathcal{B}_{b[c]}}   $          &  $5.619[2.4]  $ $GeV$       \\
$ m_{B[D]}   $          &  $5.279[1.870]  $ $GeV$       \\
$ g_{NB[D]\mathcal{B}}   $          &  $6.74  $ $GeV$       \\
$ f_{B[D]}   $          &  $0.17[0.2067]  $   $GeV$      \\
$ \rho_{N}^{sat}     $          &  $(0.11)^3  $ $GeV^3$        \\
$ \langle q^{\dag}q\rangle_{N}    $          &  $\frac{3}{2}(2 M_{N})$         \\
%$ \langle\bar{q}q\rangle_{0}           $          &  $ (-0.241)^3    $ $GeV^3$          \\
$ m_{q}      $          &  $6~MeV$                 \\
$ \sigma_{N}            $          &  $0.045   $GeV$ $                  \\
$  \langle\bar{q}q\rangle_{N}  $          &  $ \frac{\sigma_{N}}{2m_{q}}(2 M_{N})$                  \\
$  \langle q^{\dag}g_{s}\sigma
Gq\rangle_{N}  $          &  $ -0.33~GeV^2 (2 M_{N})$                  \\
$  \langle q^{\dag}iD_{0}q\rangle_{N}  $          &  $0.18 ~GeV (2 M_{N})$                  \\
$  \langle\bar{q}iD_{0}q\rangle_{N}  $          &  $\simeq0 $                  \\
%$  m_{0}^{2}  $          &  $ 0.8~GeV^2$                  \\
%$   \langle\bar{q}g_{s}\sigma Gq\rangle_{0} $          &  $m_{0}^{2}\langle\bar{q}q\rangle_{0} $                  \\
$  \langle\bar{q}g_{s}\sigma
Gq\rangle_{N}  $          &  $3~GeV^2(2 M_{N}) $                  \\
$ \langle  \bar{q}iD_{0}iD_{0}q\rangle_{N} $          &  $ 0.3~GeV^2(2 M_{N})-\frac{1}{8}\langle\bar{q}g_{s}
\sigma
Gq\rangle_{N}$                  \\
$  \langle
q^{\dag}iD_{0}iD_{0}q\rangle_{N}  $          &  $0.031~GeV^2(2 M_{N})-\frac{1}{12}\langle
q^{\dag}g_{s}
\sigma Gq\rangle_{N} $                  \\
%$\langle \frac{\alpha_s}{\pi} G^{2}\rangle_{0}$ & $(0.33\pm0.04)^4~GeV^4$\\
$\langle \frac{\alpha_s}{\pi} G^{2}\rangle_{N}$ & $-0.65~GeV (2 M_{N})$\\
 \hline \hline
\end{tabular}
\caption{Numerical values for input parameters \cite{Hayashigaki,Cohen,Nielsen,XJ1,Cohen45,Wang2011,Navarra,pdg}. $\rho_{N}^{sat}$ means the saturation nuclear matter density. }
\end{table}

%\begin{figure}[h]
%\centering
%\begin{tabular}{cc}
%\includegraphics[totalheight=6cm,width=7cm]{BMesonMassShiftPoleVersusMsqGraph.eps}
%\includegraphics[totalheight=6cm,width=7cm]{DMesonMassShiftPoleVersusMsqGraph.eps}
%\end{tabular}
%\caption{The $B$ meson mass shift in nuclear matter  versus Borel mass $M^2$ at three different values of
%continuum threshold using pole masses of heavy quarks (left
%panel). The same, but for $D$ meson mass shift  (right panel). }
%\end{figure}

%\begin{figure}[h]
%\centering
%\begin{tabular}{cc}
%\includegraphics[totalheight=6cm,width=7cm]{BMesonfBShiftPoleVersusMsqGraph.eps}
%\includegraphics[totalheight=6cm,width=7cm]{DMesonfDShiftPoleVersusMsqGraph.eps}
%\end{tabular}
%\caption{The shift of $B$ meson's decay constant in nuclear matter  versus Borel mass $M^2$ at three different values of
%continuum threshold using pole masses of heavy quarks (left
%panel). The same, but for shift in decay constant of the $D$ meson  (right panel).  }
%\end{figure}

\begin{figure}[h]
\centering
\begin{tabular}{cc}
\includegraphics[totalheight=6cm,width=7cm]{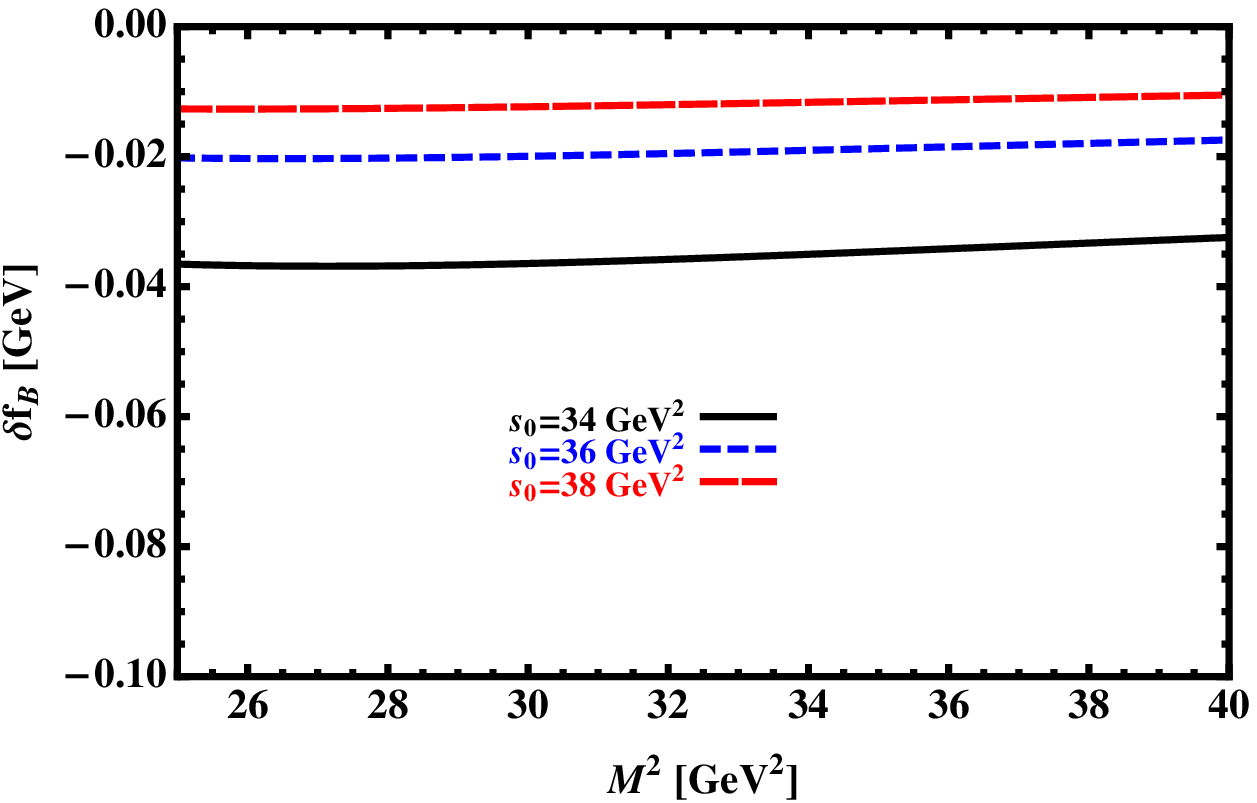}
\includegraphics[totalheight=6cm,width=7cm]{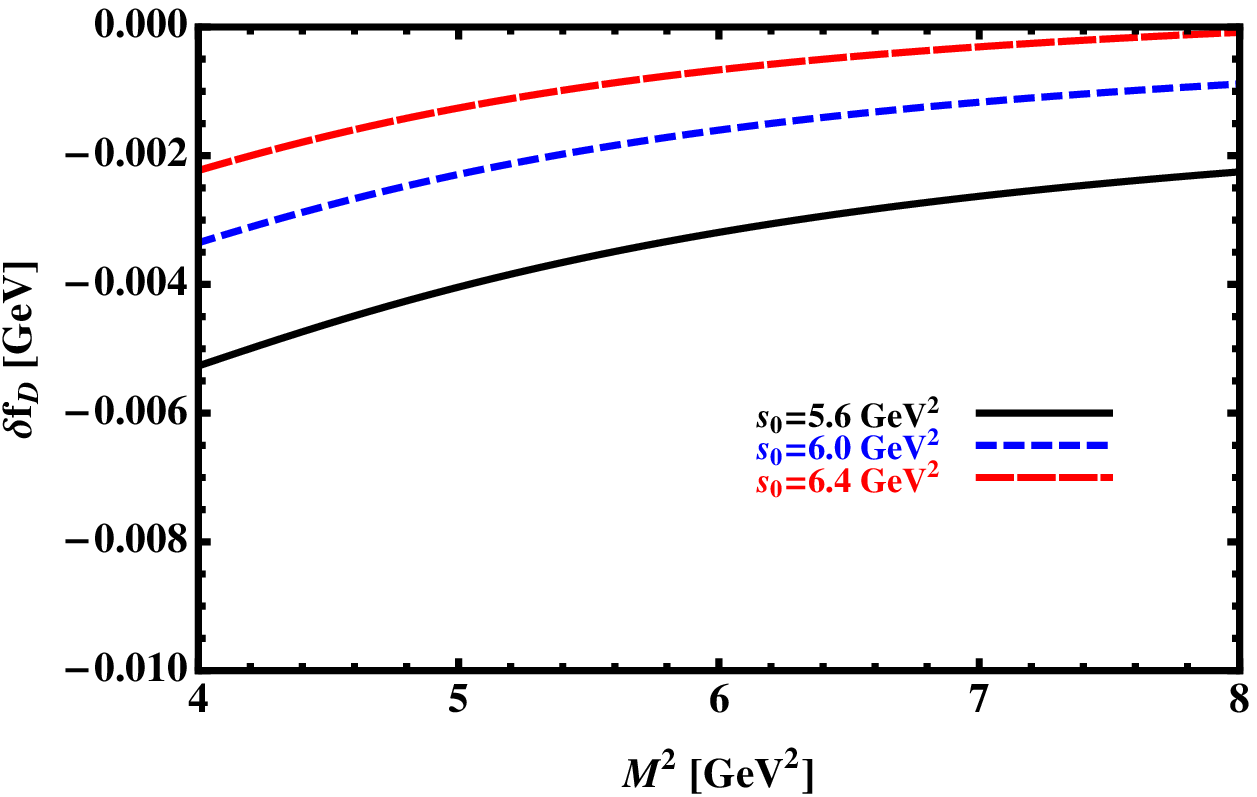}
\end{tabular}
\caption{The shift of $B$ meson's decay constant in nuclear matter  versus Borel mass $M^2$ at three different values of
continuum threshold  (left
panel). The same, but for shift in decay constant of the $D$ meson  (right panel). }
\end{figure}

\begin{table}[htdp]
\begin{center}\begin{tabular}{|l|c|c|}\hline &   $ \delta f_{B}$ (GeV)  & $ \delta f_{D}$ (GeV)  \\\hline 
%PW (pole) & $-0.310\pm0.080$ &$-0.028\pm0.009$ & $-0.066 \pm0.010$&$-0.003\pm0.002$\\\hline
Present Work   & $-0.023\pm0.007$ &$-0.002\pm0.001$   \\\hline
%\cite{Hayashigaki} &  $-$ & $-0.048\pm 0.008$ & $-$\\\hline
%\cite{Hilger} & -0.130 & - & -0.060 & - \\\hline
\end{tabular} \caption{Average values of the shifts in the  decay constants of the $B$ and $D$ mesons. }
\end{center}
\label{result}
\end{table}

Making use of all input parameters and working regions for the auxiliary parameters we depict the dependence of the shifts in the  decay constants of the $B$ and $D$ mesons on the Borel mass
parameter at different fixed values of the continuum threshold in figure 1. The left panel in this  figure belongs to the shift of the  decay constant of  the $B$ meson, while the right panel includes the
 variations of this quantity with respect to $M^2$ in $D$ channel. 
%The quantities without over-line are obtained using the pole masses of the heavy quarks, while masses and decay constants with over-line are extracted using
%the $\overline{MS}$ scheme of the heavy quark masses. 
By a quick glance at this figure, we see that 
\begin{itemize}
 %\item the masses demonstrate good stabilities with respect to $M^2$ in its working regions for both the pole and $\overline{MS}$ quark masses.
 \item the decay constant at $B$ channel  shows a good stability with respect to the variations of $M^2$ in its working region, while this quantity  weakly depend on $M^2$ at $D$ channel. 
The absolute value of the shift in the decay constant of the $D$ meson decreases by  increasing the value of $M^2$, such that at upper band of Borel mass parameter and higher value of continuum threshold this
 shift becomes very small.
\item The shifts of decay constants due to nuclear medium are negative in both  $B$ and $D$ channels.
\item The shift in the decay constant of $B$ meson is roughly 10 times bigger than that of the $D$ meson.
\item Increasing the value of the continuum threshold in both channels ends up with decrease in the absolute values of the shifts in decay constants.
\end{itemize}
Extracted from  figure 1, we depict the average values of the shifts in the decay constants of $B$ and $D$ mesons in table 2. The quoted errors in the values of the shifts in decay constants
belong to the uncertainties coming from both the determination of the working region for auxiliary parameters and errors of other input parameters.

\begin{figure}[h]
\centering
\begin{tabular}{cc}
\includegraphics[totalheight=6cm,width=7cm]{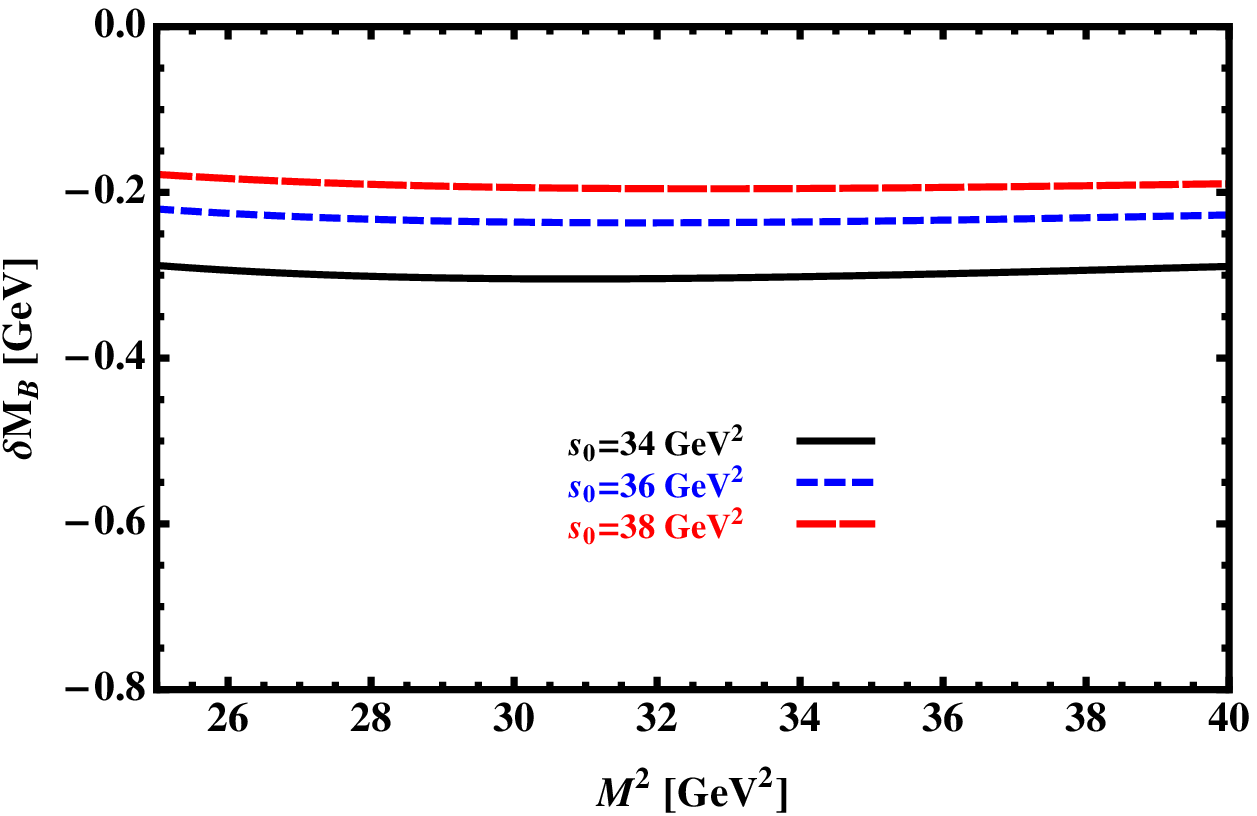}
\includegraphics[totalheight=6cm,width=7cm]{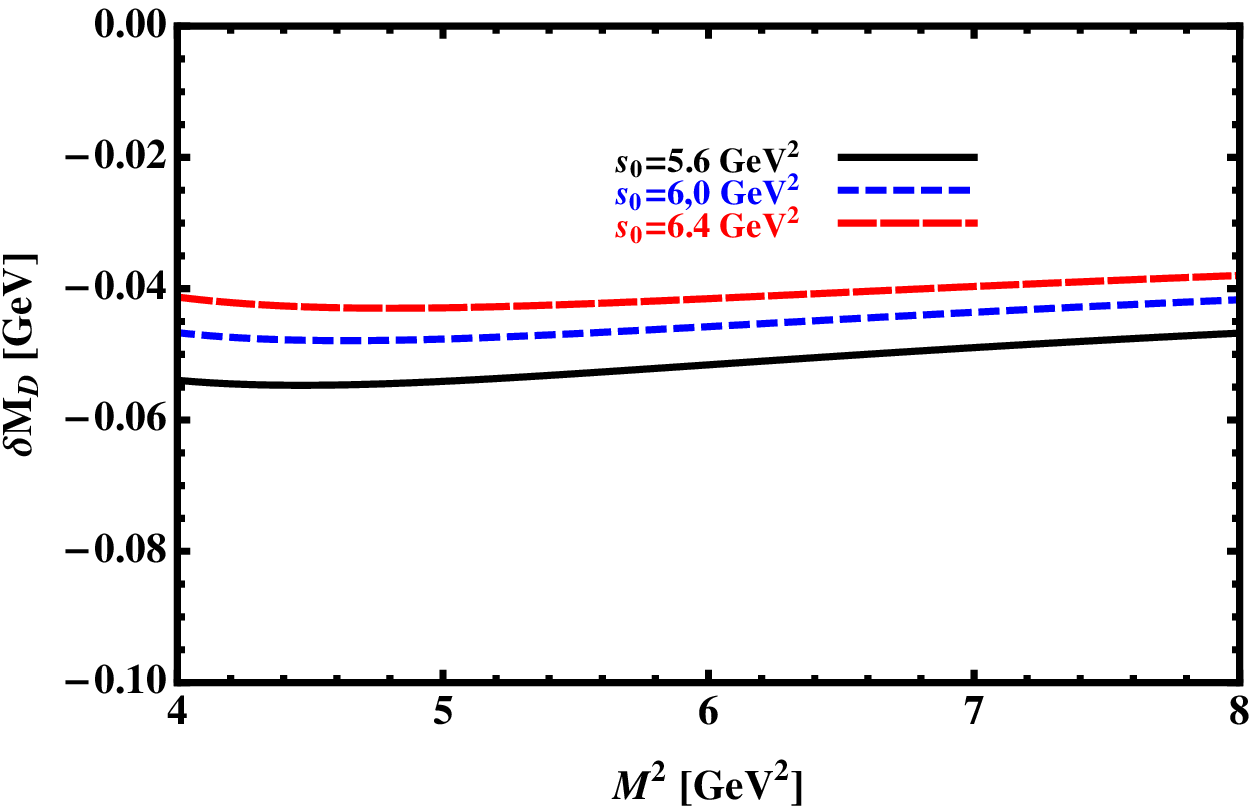}
\end{tabular}
\caption{The $B$ meson mass shift in nuclear matter  versus Borel mass $M^2$ at three different values of
continuum threshold (left
panel). The same, but for $D$ meson mass shift  (right panel). }
\end{figure}

In order to make a comparison of the results on the mass shifts with the previous theoretical predictions, we also numerically analyze these shifts in the $B$ and $D$ channels. 
For this aim, we depict the dependence of
the mass  shifts on $M^2$ at different fixed values of $s_0$ in figure 2. From this figure, we conclude that
\begin{itemize}
 \item the shifts in the masses  of both $B$ and $D$ mesons are negative.
\item The shift in the mass of the $B$ meson is roughly 5 times grater than that of the $D$ meson.
\item The shifts in both $B$ and $D$ channels demonstrate good stabilities with respect to the variations of the Borel mass parameter. 
\end{itemize}

From figure 2, we also extract the values of the shifts in the masses of the mesons under consideration as presented in table 3. For comparison, we also depict the  predictions of some previous theoretical works
in the same table. From this table, we see that  our result on the mass shift in   $D$ channel is in a good consistency with the result of  \cite{Hayashigaki} which uses the same method and interpolating current. 
However, the prediction of \cite{Hilger} in this channel is in opposite sign with ours and prediction of \cite{Hayashigaki}, although it predicts the same value  in magnitude.  As far as the shift in the mass of 
 $B$ channel is considered, the only existing prediction belongs to \cite{Hilger} which is different than our result in both sign and magnitude. Note that in \cite{Hilger} the authors use the interpolating currents
$J_{D^+}=i\bar d \gamma_5 c$ and $J_{B^+}=i\bar b \gamma_5 u $ or $J_{B^0}=i\bar b \gamma_5 d $ in $D$ and $B$ channels, respectively.
%with the same method and same interpolating current for the pseudoscalar mesons on the shift in the mass of the $D$ meson \cite{Hayashigaki} in the same table. From this table we conclude that
%\begin{itemize}
% \item the shifts in the masses and decay constants of both $B$ and $D$ mesons are negative.
 %\item The shift in the mass of the $B$ meson is roughly 5 times grater than that of the $D$ meson.
 %\item We see a considerable negative shift in the decay constant of the $B$ meson, while this shift in the case of $D$ meson is very small.
%\item The results have considerable dependence on the heavy quark masses, such that when switching from the pole masses to  $\overline{MS}$ masses we see considerable decreases in the absolute values of the 
%shifts in the quantities under consideration.
%\item Our result on $ \delta m_D$ is comparable with the prediction of \cite{Hayashigaki} when the $\overline{MS}$ mass of the $c$ quark is used, however, our prediction is considerably high when the pole mass of the charm quark
%is used.
%\end{itemize}

\begin{table}[htdp]
\begin{center}\begin{tabular}{|l|c|c|}\hline & $\delta m_{B}$ (GeV)  & $ \delta m_D$ (GeV)  \\\hline 
%PW (pole) & $-0.310\pm0.080$ &$-0.028\pm0.009$ & $-0.066 \pm0.010$&$-0.003\pm0.002$\\\hline
Present Work  & $-0.242\pm0.062$ & $-0.046\pm0.007$  \\\hline
\cite{Hayashigaki} & $-$ &  $-0.048\pm 0.008$ \\\hline
\cite{Hilger} & $\sim0.060$ &  $\sim 0.045 $ \\\hline

%\cite{Hilger} & -0.130 & - & -0.060 & - \\\hline
\end{tabular} \caption{Average values of the shifts in the masses  of the $D$ and $B$ mesons.  }
\end{center}
\label{result}
\end{table}

At the end of this section, we would like to discuss the dependence of the results on the shifts in the decay constants and masses to the nuclear matter density. In the above numerical results,
 we have  used the value of saturation density, i.e. $\rho_N^{sat}=(0.11)^3 ~GeV^3$. In order to see how the results depend on the nuclear matter density, we plot the shifts in
 the decay constants and masses versus $\rho_N/\rho_N^{sat}$ in figures 3 and 4 at the average values of the Borel mass parameter and continuum threshold.
\begin{figure}[h]
\centering
\begin{tabular}{cc}
\includegraphics[totalheight=6cm,width=7cm]{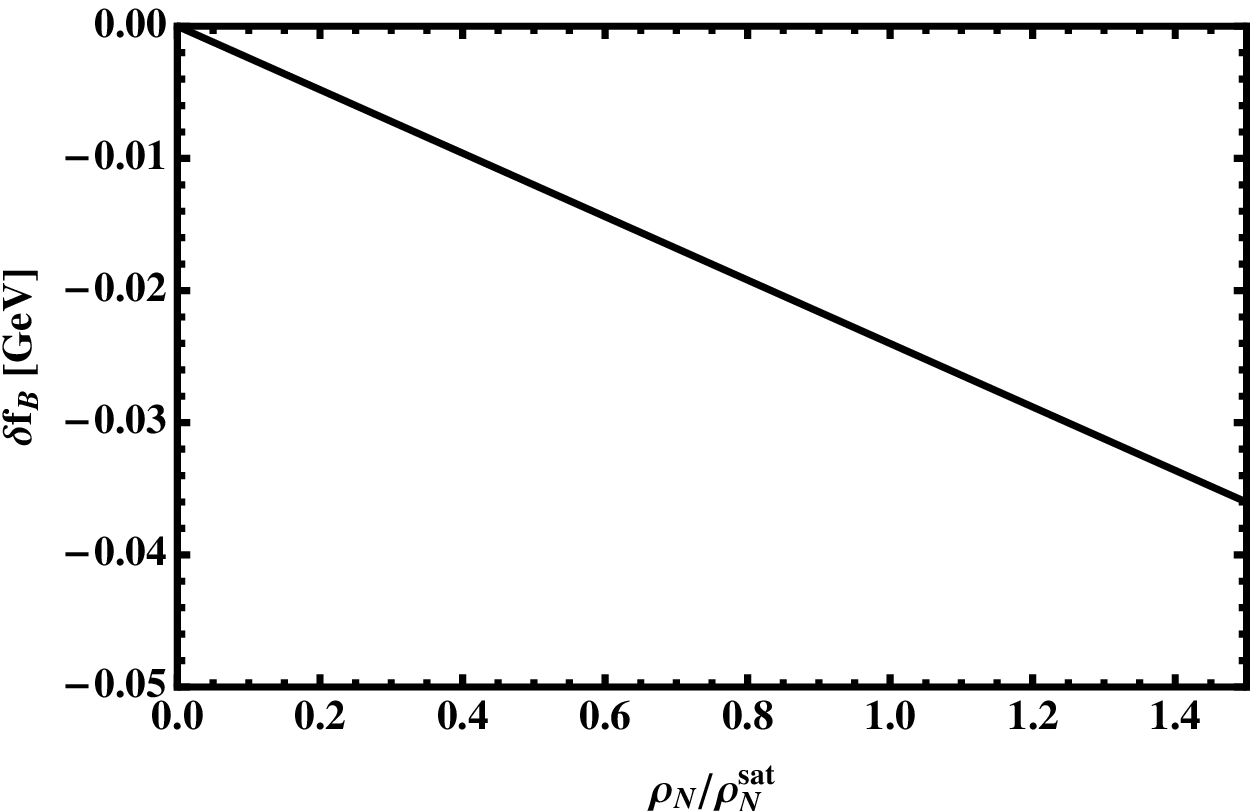}
\includegraphics[totalheight=6cm,width=7cm]{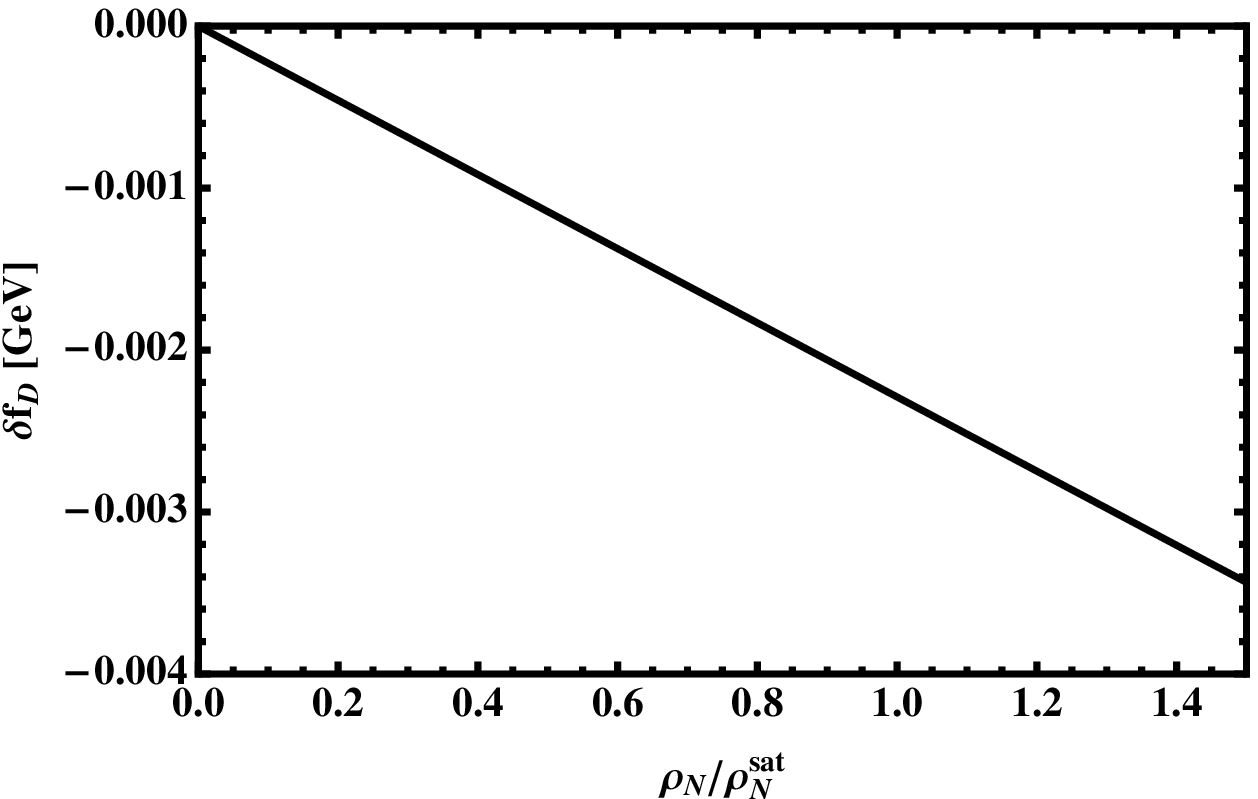}
\end{tabular}
\caption{The dependence of the  shift in the decay constant of the $B$  meson to the  nuclear matter  density (left
panel). The same, but for $D$ meson  (right panel). }
\end{figure}
\begin{figure}[h]
\centering
\begin{tabular}{cc}
\includegraphics[totalheight=6cm,width=7cm]{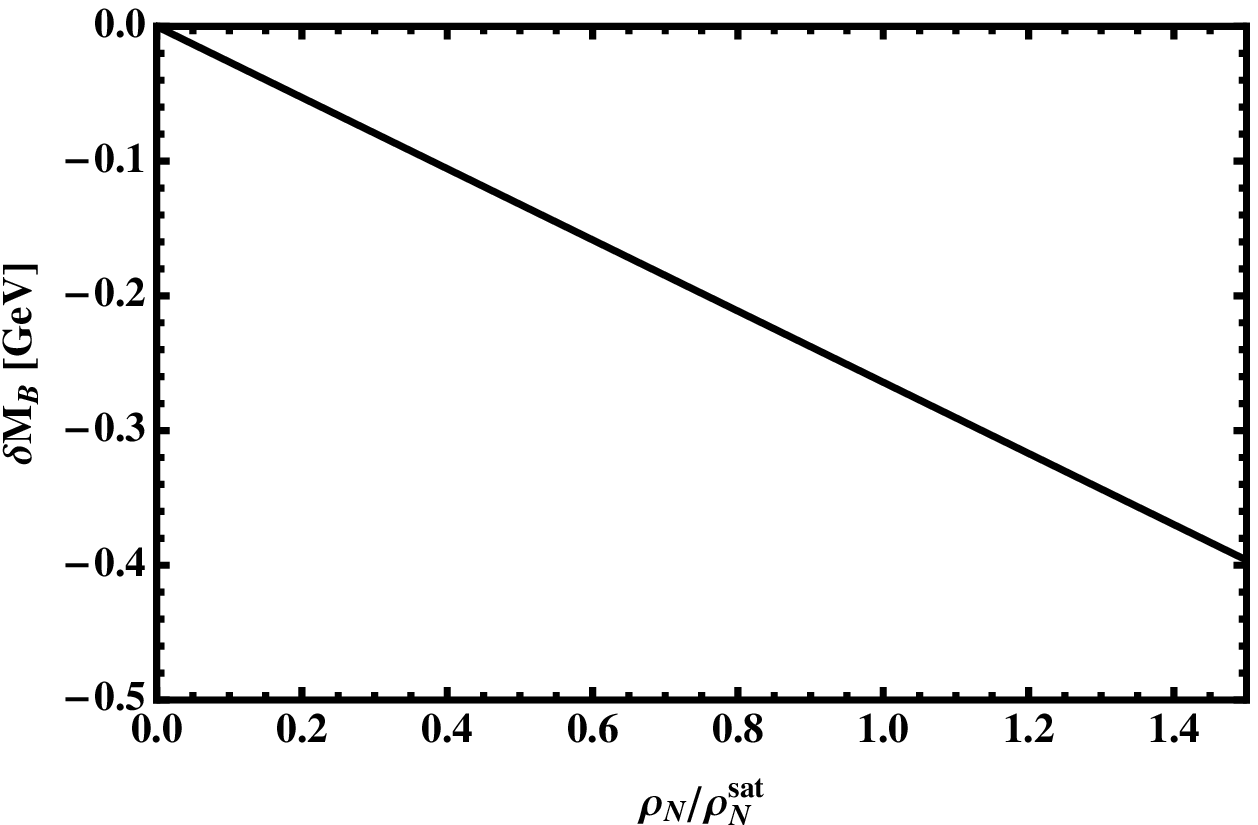}
\includegraphics[totalheight=6cm,width=7cm]{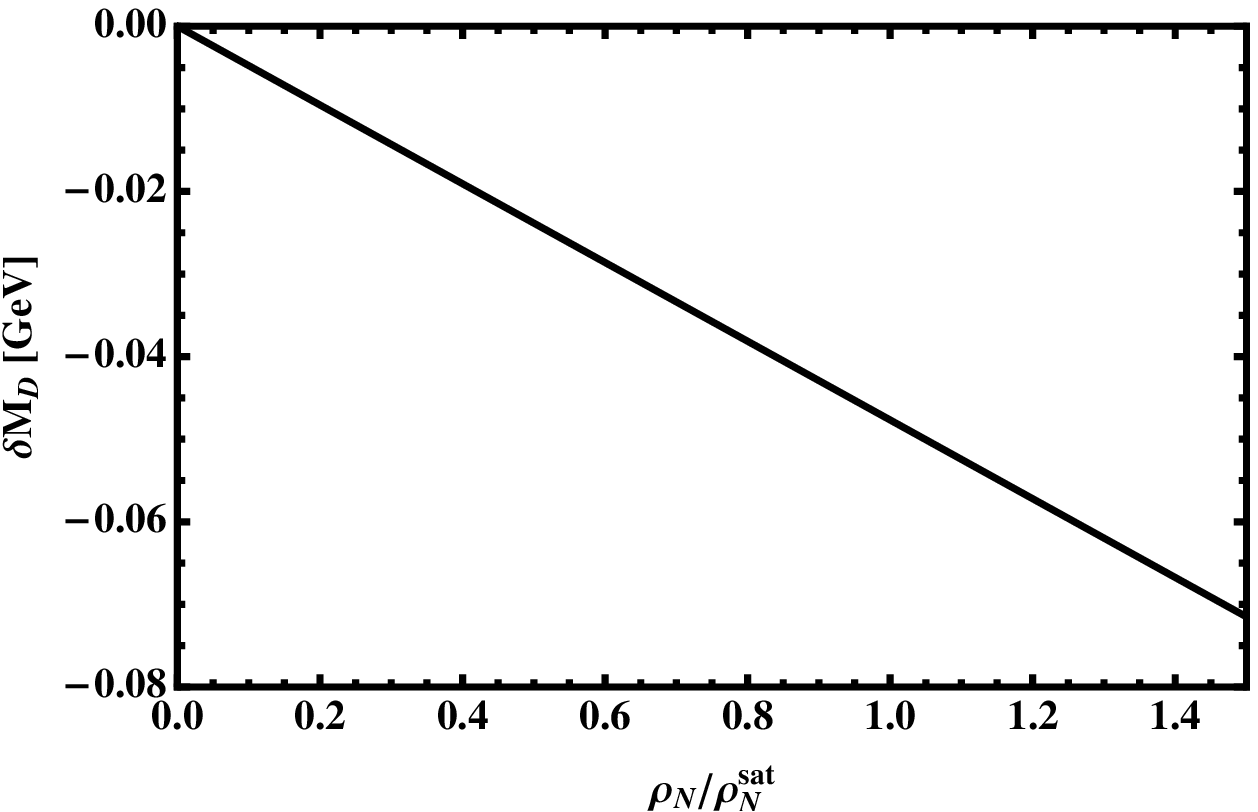}
\end{tabular}
\caption{The dependence of the  shift in the mass of the $B$  meson to the  nuclear matter  density (left
panel). The same, but for $D$ meson  (right panel). }
\end{figure}
As also expected from Eqs. (\ref{massshift}) and (\ref{lepdecconst}),  these figures show that the shifts in the physical quantities under consideration linearly depend on the nuclear matter density.
 The absolute values of the shifts in the decay constants and masses increase by
increasing the nuclear matter density.

In summary, we calculated the shifts in the decay constants  and masses of the pseudoscalar $D$ and $B$ mesons due to nuclear matter  in the framework of the QCD sum rules. We found considerable negative shifts in the values of 
the considered quantities  except for the shift in the decay constant of the  $D$ meson which is very small. We  compared our results on the mass shifts in $D$ and $B$ channels with the predictions
of some existing theoretical works in the literature. We also discussed the dependence of the shifts in the decay constants and masses of these mesons on the nuclear matter density.  The results obtained in the present work can be useful in analyzing  the future experimental data at
 different heavy ion
collision experiments. The results obtained  for the shift in masses especially for those in the decay constants  can also be used in theoretical calculations of the electromagnetic
 properties of the considered mesons as well as their strong couplings with other hadrons in nuclear medium.

\section{Acknowledgment}
This work has been supported in part by the Scientific and Technological
Research Council of Turkey (TUBITAK) under the research project 114F018.

%\newpage
 
                                 %%%%%%%%%%%%%%%%%%%%%%%%%%%%%%%%%%%%%%%
       %%%%%%%%%%%%%%%%%%%%%%%%%%%%%%%%%%%%%%%             %%%%%%%%%%%%%%%%%%%%%%%%%%%%%%%%%%%%%%%%%%
                                 %%%%%%%%%%%%%%%%%%%%%%%%%%%%%%%%%%%%%%%
\end{document}